\documentclass[journal,twocolumn]{IEEEtran}
\usepackage{graphicx}
\usepackage{amsmath}
\usepackage{subfigure}
\usepackage{multirow}
\usepackage{algorithmic}
\usepackage[linesnumbered,ruled,vlined]{algorithm2e}

\begin{document}

\title{Stand-alone device for IoT applications}

\author{Neha K Nawandar, Naveen Cheggoju and Vishal R Satpute}

\maketitle

\begin{abstract}
Internet of Things (IoT) is a digital world of connected and talking devices, providing room for countless and diverse smart applications. This paper proposes one such IoT enabled stand-alone device with numerous capabilities: (i) interaction with user, (ii) required application selection, (iii) data sensing, (iv) data publish, and (v) decision making and actuation. The algorithm allows user to pick an application and input specific data for calibration, which on completion enables the device for further working. To verify and test its capability, a smart home garden environment is created using this device, temperature, humidity, soil moisture sensors and actuators. As it is implicit that real-time communication is inevitable for an IoT application, the sensor data is published to a Mosquitto MQTT broker to permit real-time remote access. The decision taken by the device is sent to actuators via relay, thus a continuous monitoring process is achieved. Results are obtained for the application which proves the device suitability for IoT applications. 
\end{abstract}


\IEEEpeerreviewmaketitle

\section{Introduction}
The rapid growth in digitization has made Internet a basic necessity and made the world digital, where people have limited time and less accuracy when compared to machines. Only if we have systems that know everything about different things based on the gathered data without any human assistance, life could be much easy. It would become very simple and easy to keep a trace down, count everything and reduce any waste, loss and cost to much an extent, we would know when a thing needs to be repaired, replaced, etc. A future evaluation of Internet, the Internet of Things (IoT) \cite{ashton2009internet} can be referred to as a system of inter-related computing devices, objects, things, animals, people, etc. that possess the capability to share information over a network without human involvement. Internet has been expanding everyday which gives rise to new opportunities for creating a link in between real physical world and the virtual world and this collaboration makes both the computing and networking more efficient, adaptable, secure and reliable for e-planes, vehicle collision avoidance, in WSNs, device monitoring, e-health and such other applications. 

The evolution of IoT has arisen from the convergence of various micro-electromechanical systems, micro-services, wireless technologies and the Internet that has helped to allow random and unstructured data generated by the machines to be analyzed for driving any improvements. IoT comprises of sensors, communication organization, processing units, decision making algorithms and the action invoke unit. Each object is unique, can be accessed over the Internet. The sensors get information and communicate it over Internet to the processing unit. Finally the processing unit output is passed to the decision making algorithm for further action. Internet growth has boosted technical power of the devices in user hand and it has connected billions of such objects all over the globe. The devices differ in dimensions, processing capabilities, computational complexities and support distinct applications. This pushes the conventional Internet to amalgamate with sophisticated and smart IoT. It has the potential to connect devices, implant intelligence in the designed systems to make them capable of processing the sensor information smartly, take practical and neat decisions  without human involvement. IoT can thus give rise to a wide range of useful and competent applications that humans had never envisioned before. 

It gives an opportunity to establish a wide range of applications and many of such have been successfully put into use. A few include wearables, smart grid, smart infrastructure, connected car, smart retail, smart supply chain, industrial internet, etc. These applications have been explored and some existing IoT products  are mentioned in Table \ref{table:iotproducts}.

\begin{table}[!h]
 \centering
 \caption{IoT products}
 \begin{tabular}{|l|l|}
 \hline
 \textbf{Application} & \textbf{Product} \\
 \hline
 Home safety & Birdi \\
 \hline
 Connected Car & Cloud Your Car \\
 \hline
 Home Ordering Tool & Amazon Dash Button \\
 \hline
 Connected Garden & Parrot Pot \\
 \hline
 Thing Tracker & Tile \\
 \hline
 Medical Alert System & Lively Watch \\
 \hline
 Pet Activity Tracker & FitBark \\
 \hline
 \end{tabular}
 \label{table:iotproducts}
\end{table}

These things communicate with each other and with the server using various protocols at different layers as given below, 
\begin{itemize}
 \item Infrastructure: 6LowPAN, IPv4/IPv6, RPL.
 \item Identification: EPC, uCode, IPv6, URIs.
 \item Communication: Wifi, Bluetooth, LPWAN.
 \item Discovery: Physical Web, mDNS, DNS-SD.
 \item Data protocols: MQTT, CoAP, AMQP, Websocket, Node.
 \item Device management: TR-069, OMA-DM.
 \item Semantic: JSON-LD, Web Thing Model
 \item Multi-layer frameworks: Alljoyn, IoTivity, Weave, Homekit.
\end{itemize}

Now being called the language of the IoT, it seems obvious that any standard that is aiming to bring a common network service layer to the IoT architecture should be able to utilize MQ Telemetry Transport (MQTT) \cite{velez2018ieee}. In this paper the device-server communication is done using the MQTT protocol due to its simplicity and low latency for data transfer. The paper is organized as: Section \ref{section:proposeddevice} introduces the stand alone device, explains its algorithm and explores it capabilities, while its results for a home gardening application have been shown in Section \ref{section:application}. Finally the paper has been concluded in Section \ref{section:conclusion} and references have been mentioned at the end.

\section{Proposed stand-alone device} \label{section:proposeddevice}
\begin{figure}[!t]
 \centering
 \includegraphics[width=0.8\columnwidth]{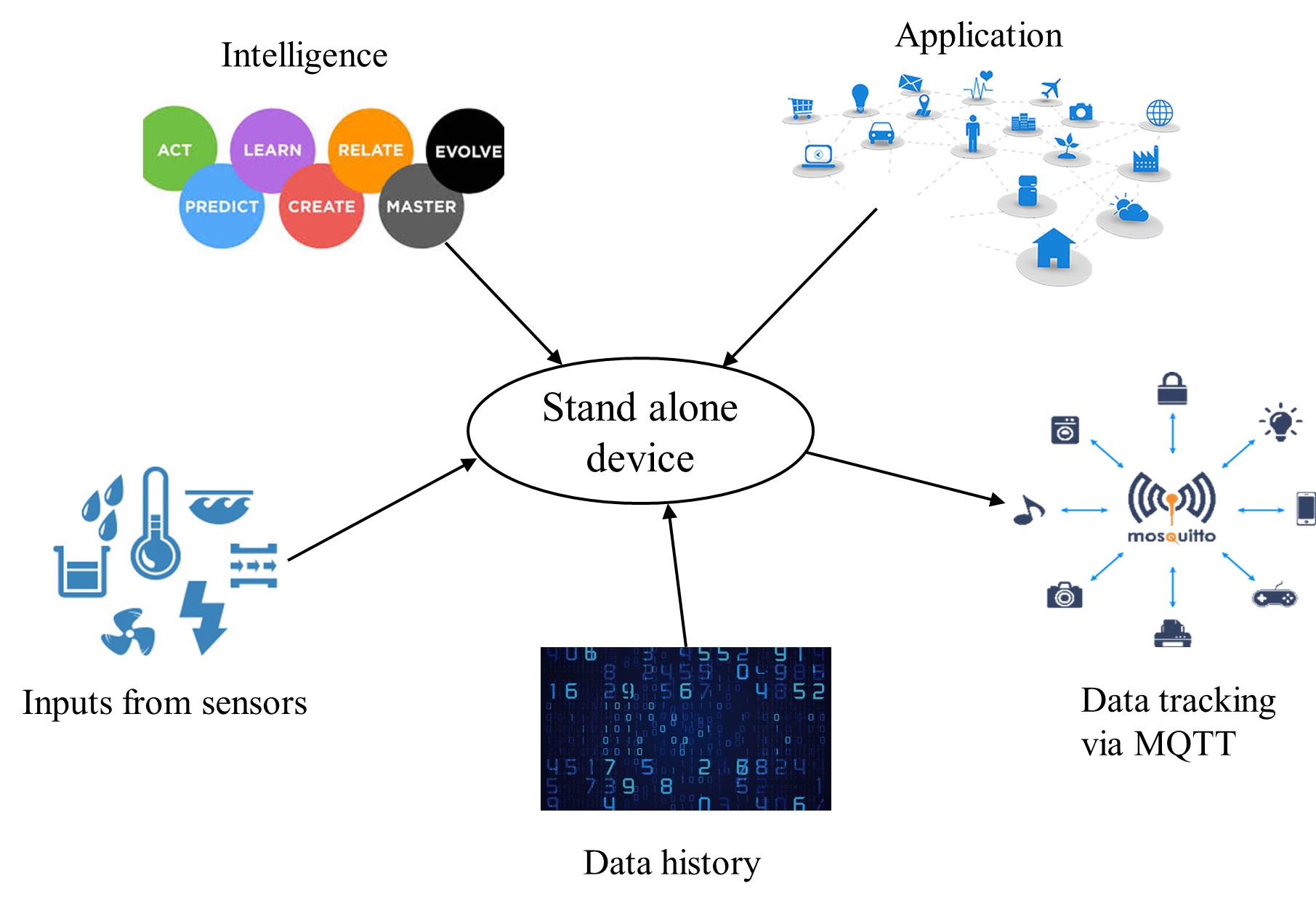}
 \caption{Proposed device capabilities}
 \label{figure:devicearch}
\end{figure}

\begin{figure}[!t]
 \centering
 \includegraphics[height=2.2cm,width=0.7\columnwidth]{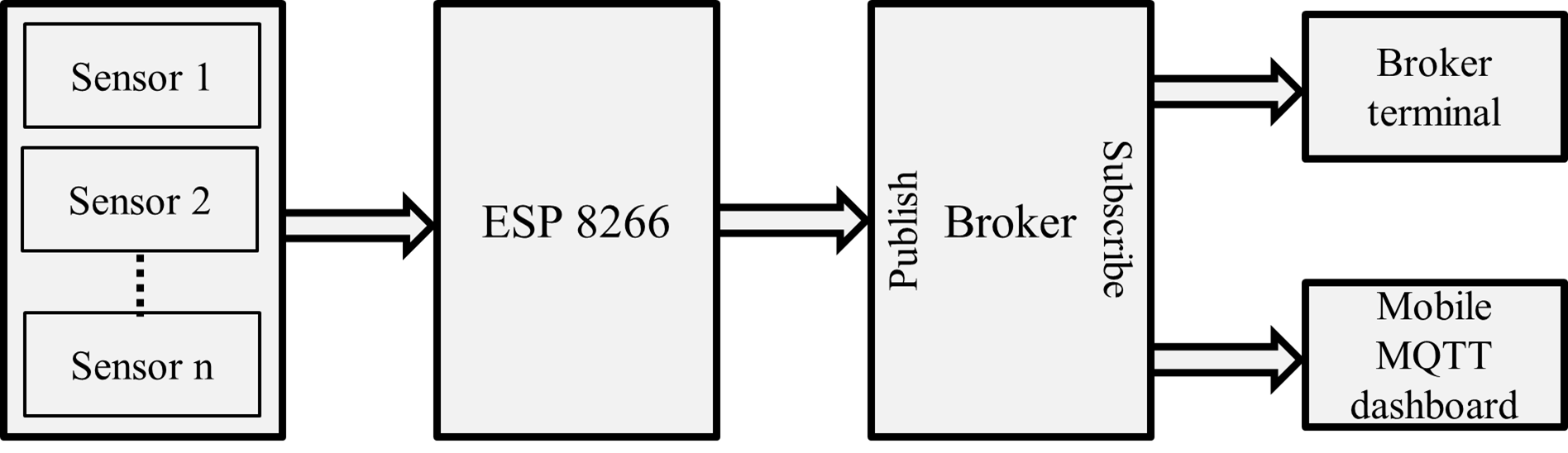}
 \caption{System architecture using proposed device}
 \label{figure:sysarch}
\end{figure}

This section introduces the proposed stand-alone device and the algorithm. The device capabilities as seen from Figure \ref{figure:devicearch} has intelligence using which it gets data from the sensors, uses current data and previous data history to predict and act accordingly for numerous applications. It also allows tracking of sensor data using MQTT protocol. The system architecture using proposed device is shown in Figure \ref{figure:sysarch} and the device algorithm in given by Algorithm \ref{algorithm:overall}. Figure \ref{figure:sysarch} shows that multiple sensors are connected to the device ESP8266 which operates as per proposed Algorithm \ref{algorithm:overall} and the data is published to the broker which can be remotely observed on mobile, dashboard, etc. The device runs an html webpage where the user can provide inputs as per the application environment and it has sensing and decision making capabilities using sensors and actuators respectively. The GPIO pins on the device connects multiple sensors and actuators as per requirement where application specific calibration can be done. The algorithm on the device runs multiple routines that are responsible for its efficient working and its pseudo code can be seen by Algorithm \ref{algorithm:overall}. Here, $\overline{S}$: set of sensors for fetching physical data, $\delta$: application specific decision, $\overline{\alpha}$: actuators required. 

The algorithm first runs in a user-device communication mode for a pre-defined time interval. In this mode the device takes inputs from the user if any and saves it temporarily onto the device for further usage. After time-out it disables this mode and starts working normally where necessary computations are done in a one time device setup routine. After successful calibration and setup, the device then starts sensing the data using the attached sensors and continuously looks for actuation if needed. It also enables the user to track down and monitor the sensor data anytime and from anywhere. This is achieved by uploading the data to a MQTT server using which the user is aware of the present scenario and can manually control the actuators if needed. 

\begin{algorithm}
\SetAlgoLined
\caption{Proposed stand-alone device pseudo-code}
\label{algorithm:overall}
\KwIn{$\overline{S}$}
\KwOut{$\delta$, $\overline{\alpha}$}
User-device communication mode:
\hspace{400pt} $n_{config}$ $\rightarrow$ n/w configuration;
\hspace{400pt} $n_{info}$ $\rightarrow$ n/w information;
\hspace{400pt} $application$:
\hspace{600pt} $user$ $input$;
\hspace{600pt} $save$ $data$ $on$ $device$;\\
\lIf{!Time-out}{Goto 1}
\Else
{
Disable current mode; \\
One time device setup; \\
Monitor $\overline{S}$: 
\hspace{600pt} $T$ $\rightarrow$ ambient temperature;
\hspace{400pt} $H$ $\rightarrow$ ambient humidity;
\hspace{400pt} $SM$ $\rightarrow$ soil moisture \\
Connect to MQTT server: 
\hspace{400pt} $Mosquitto$ $broker$ $\rightarrow$ $10.4.1.100$;
\hspace{400pt} $Create$ $topics$ $\rightarrow$ /usp/temp, /usp/humid, /usp/sm;
\hspace{400pt} $Publish$ $data$; \\
Save data offline; \\
Alert to relay $\rightarrow$ control $\overline{\alpha}$
}
\end{algorithm}


\begin{table}[!t]
 \centering
 \caption{Smart application sensor requirements}
  \begin{tabular}{|p{2.8cm}|p{5cm}|}
 \hline
  Application & Sensors \\
  \hline
  \multirow{5}{*}{Smart home \cite{sequeiros2021impact}} & Temperature \\ \cline{2-2}
  & Humidity  \\ \cline{2-2}
  & Light intensity  \\ \cline{2-2}
  & Motion \\ \cline{2-2}
  & Fire/ CO detection \\
  \hline
  \multirow{5}{*}{Smart e-health \cite{scarpato2017health} } & Epidermal (Bio-patches, digital tatoos, body-worn tags) \\ \cline{2-2}
  & RFID tag medicines \\ \cline{2-2}
  & Soluble ingestible embedded in pills \\ \cline{2-2}
  & Blood sampling sensors \\ \cline{2-2}
  & Tissue embedded (Pacemakers, defibrillators) \\
  \hline
  \multirow{4}{*}{Intelligent parking \cite{said2021intelligent}} & IR \\ \cline{2-2}
  & Rear-mounted cameras \\ \cline{2-2}
  & Proximity \\  \cline{2-2}
  & RADAR \\
  \hline
  \multirow{4}{*}{Waste management \cite{anagnostopoulos2017challenges}} & PID \\ \cline{2-2}
  & Fill-level \\ \cline{2-2}
  & Temperature \\ \cline{2-2}
  & Trash sonar \\
  \hline
  \multirow{4}{*}{Smart farming \cite{zamora2019smart}} & Temperature \\ \cline{2-2}
  & Humidity \\ \cline{2-2}
  & Soil (pH, nutrients, moisture) \\  \cline{2-2}
  & Rain \\
  \hline
  \multirow{5}{*}{Env. monitoring \cite{kelly2013towards}} & CO and leakage detection \\ \cline{2-2}
  & Temperature and humidity \\ \cline{2-2}
  & Thermo-hygrometer \\ \cline{2-2}
  & Snow Level \\ \cline{2-2}
  & Earth density \\
  \hline
 \end{tabular}
 \label{table:sensorsandapplications}
\end{table}

As seen in step 7 of Algorithm \ref{algorithm:overall}, the device establishes a connection with Mosquitto, an open source MQTT broker for publishing the sensor data. MQTT being a light-weight, energy efficient and way faster than HTPP, is preferred here in order to reduce the power requirement and latency. The data is sent to this Mosquitto broker where the user continuously tracks it and statistics is also generated. If an unsuccessful connection occurs it might create data loss, so to prevent it a compressed backup is also stored offline on the device. This increases the device validity, which implies that it is capable of adapting to changes thus making it efficient for future use without the need of recalibration. The sensor data is used for actuation and so a decision $\delta$ based on the sensed data alerts the $\overline{\alpha}$ thus maintaining a continuous monitoring loop. The device connects to relays that can be used to connect to multiple application specific actuators $\overline{\alpha}$. Possible sensors for some smart applications have been mentioned in Table \ref{table:sensorsandapplications}. The proposed device can be used for these applications and the corresponding sensors when connected to the GPIO pins completes the working. 

As mentioned earlier, the device hosts an html page and the Figure \ref{figure:webpage} (a) shows three options: (i) network configuration, (ii) network information and, (iii) application, that are available in user-device communication mode. The device scans for all nearby available networks and can be configured to connect to any of these using the network configuration button. On selection, it displays the connected network and its details and, the nearby networks. The final network information is then available on the network information page. The user inputs are obtained using the application button. To test the efficiency and working of proposed device, a smart home gardening application environment is created and, $\overline{S}$, $\overline{\alpha}$ are chosen accordingly. Section \ref{section:application} introduces the application and presents the respective results obtained from the same.

\section{Application and results} \label{section:application}
The proposed stand-alone device is tested for its efficacy using a smart home gardening application and some of the sensors mentioned in smart farming portion of Table \ref{table:sensorsandapplications} are used for getting data. It is used to,
\begin{itemize}
 \item Get the ambient characteristics.
 \item Fetch soil moisture.
 \item Data publish to MQTT broker.
 \item Compute irrigation (IN) requirements.
 \item Give $\delta$ for drip.
 \item Motor control for IN.
\end{itemize}

The device is used for a smart farming application and apart from its basic capabilities, the above mentioned are achieved using it. It can be used to track the growth and determine IN needs of crops planted in backyard gardens, indoor or organic farming, vertical gardens, fields, etc. To achieve this, temperature and humidity sensor for ambient data and, soil moisture sensor for getting soil moisture level are connected to the device. In this case the user inputs taken are: crop, the type of soil to be planted in and the date of crop plantation. These data altogether are used to compute the IN requirements and an irrigation schedule. It is then connected to an actuator, a water pump in this case to supply water when required. 

\begin{figure}
 \centering
 \includegraphics[width=0.7\columnwidth]{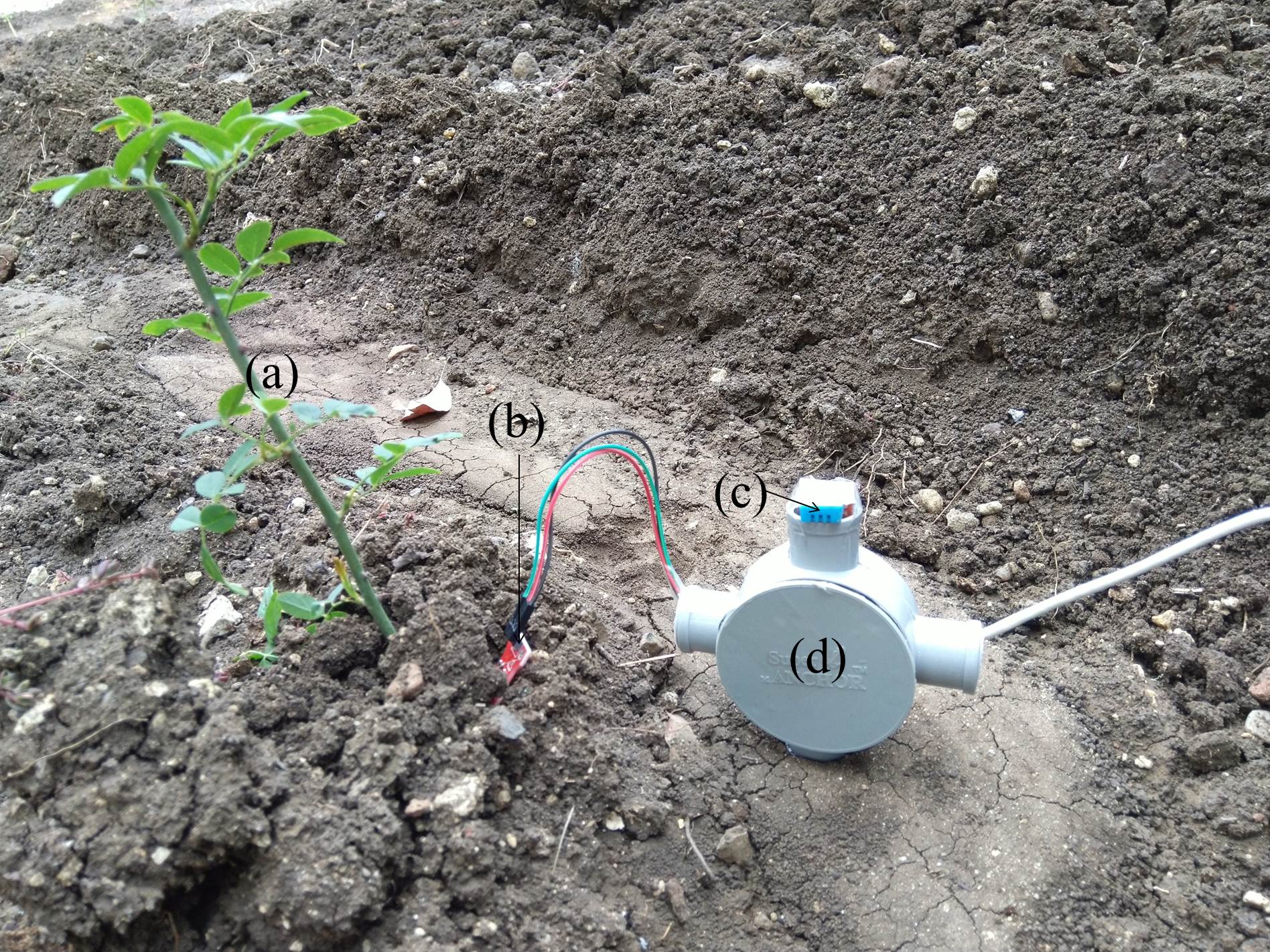}
 \caption{Application setup}
 \label{figure:applicationsetup}
\end{figure}

A setup consisting of sensors, actuators is needed for the application and the arrangement is as shown in Figure \ref{figure:applicationsetup}. Here, (a) crop unser test, (b) soil moisture sensor: gets the soil water content, (c) temperature and humidity sensor: fetches the ambient data, (d) the controller: controls the sensors, actuators and works as per Algorithm \ref{algorithm:overall}. A detailed working of this stand alone device for a smart farming irrigation based application \cite{nawandar2019iot} has been done.

\begin{figure}[!t]
 \centering
 \subfigure[Admin page]{\includegraphics[height=4.5cm,width=0.4\columnwidth]{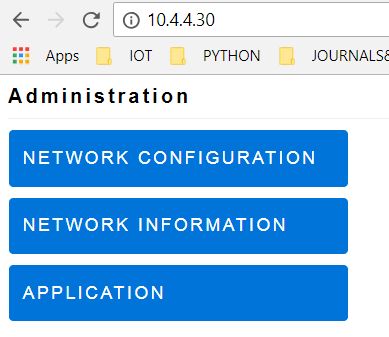}} \hspace{10pt}
 \subfigure[Network Information]{\includegraphics[height=4.5cm,width=0.4\columnwidth]{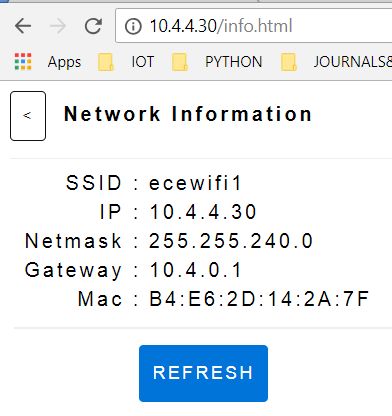}} \hspace{10pt} \\
 \subfigure[Network details]{\includegraphics[height=4.5cm,width=0.4\columnwidth]{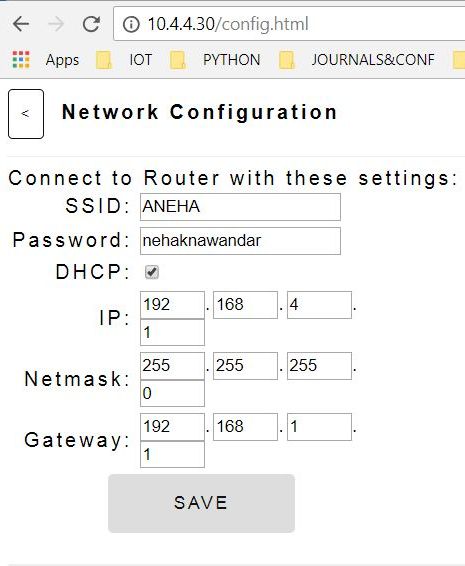}} \hspace{10pt}
 \subfigure[Available networks]{\includegraphics[height=4.5cm,width=0.4\columnwidth]{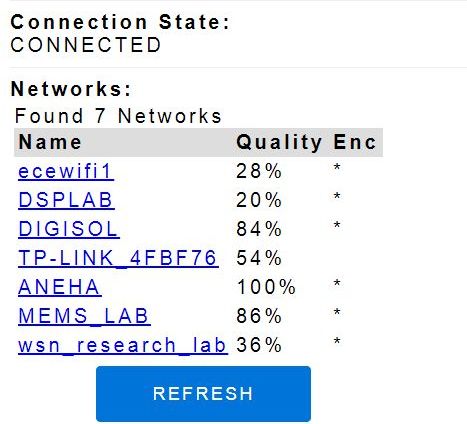}} \\
 \subfigure[Crop options]{\includegraphics[height=4.5cm,width=0.4\columnwidth]{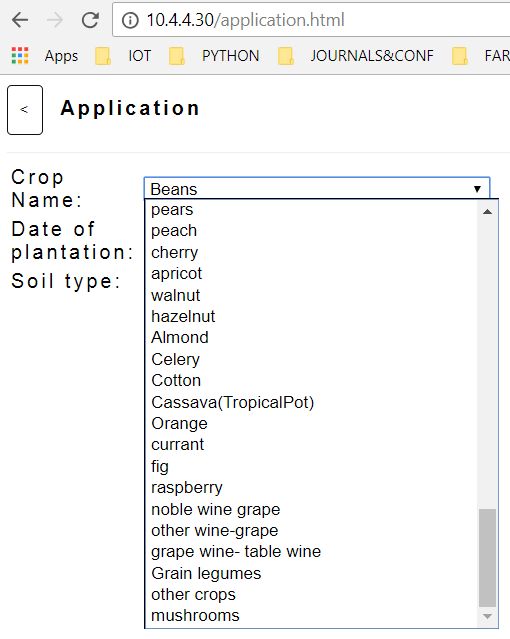}} \hspace{10pt}
 \subfigure[User inputs]{\includegraphics[height=4.5cm,width=0.4\columnwidth]{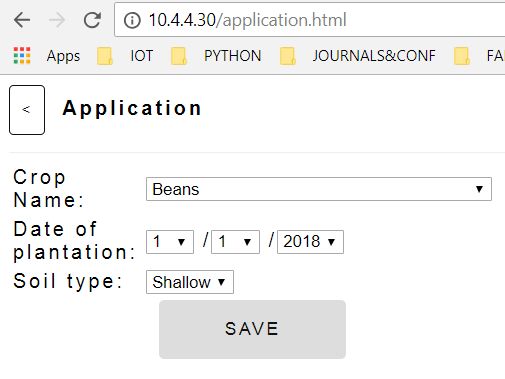}}
 \caption{User-device communication HTTP webpage}
 \label{figure:webpage}
\end{figure} 

\begin{figure}[!h]
\centering
 \subfigure[Stagewise details]{\includegraphics[height=4.6cm,width=0.43\columnwidth]{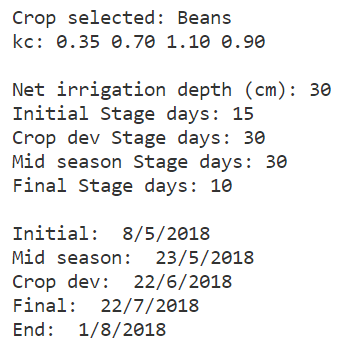}} \hspace{20pt}
 \subfigure[IN data]{\includegraphics[height=4.6cm,width=0.43\columnwidth]{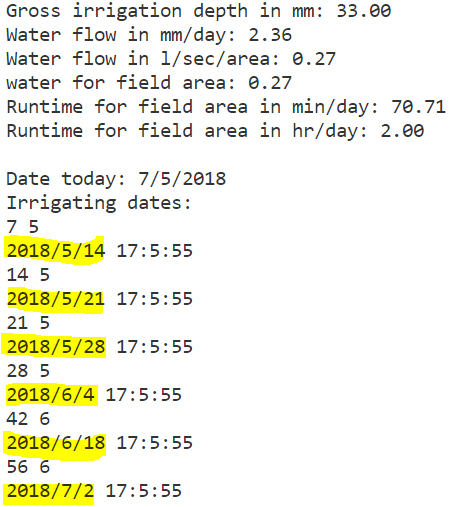}}
 \caption{Computed parameters for Beans}
 \label{figure:beans}
\end{figure}

An application option on the webpage as shown in Figure \ref{figure:webpage} (a) is available which on selection generates fields for user input consisting of crop, plantation date and soil type. A list of crops seen from Figure \ref{figure:webpage} (e) is available from which a selection is made for a crop that needs to be planted. In a similar fashion, dates and soil type are selected and these are stored onto the device for further usage as seen in Figure \ref{figure:webpage} (f). Based on these selections and a history of weather data, some crop related parameters are selected and computed that help in determining the IN needs and the IN schedule. The device is intelligent enough to tackle abrupt climatic conditions and accordingly update the IN needs. Some sample results can be seen in Figure \ref{figure:beans}. Figure \ref{figure:beans}(a) shows the number of days in every growth stage and the starting dates for the same, and Figure \ref{figure:beans}(b) gives the IN depth, water runtime and dates (highlighted). Using this data provided by the device, user can grow plants whenever needed. 


\begin{figure}[!ht]
\centering
 \subfigure[Broker status]{\includegraphics[height=4cm,width=0.35\columnwidth]{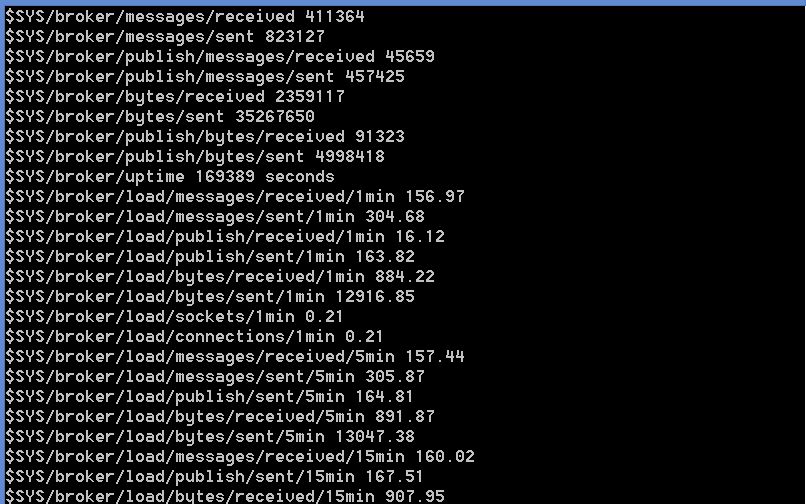}} \hspace{0.1cm}
 \subfigure[Subscribed topic on broker]{\includegraphics[height=4cm,width=0.35\columnwidth]{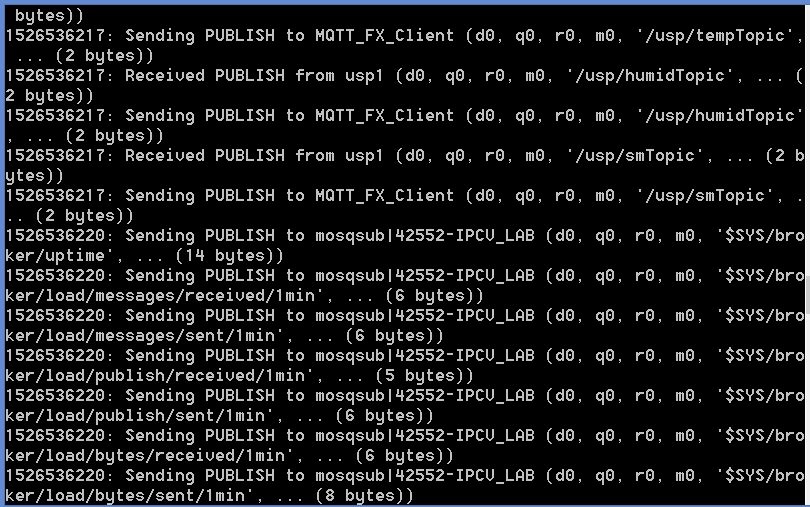}} \hspace{0.1cm}
 \subfigure[Data on dashboard]{\includegraphics[height=4cm,width=0.35\columnwidth]{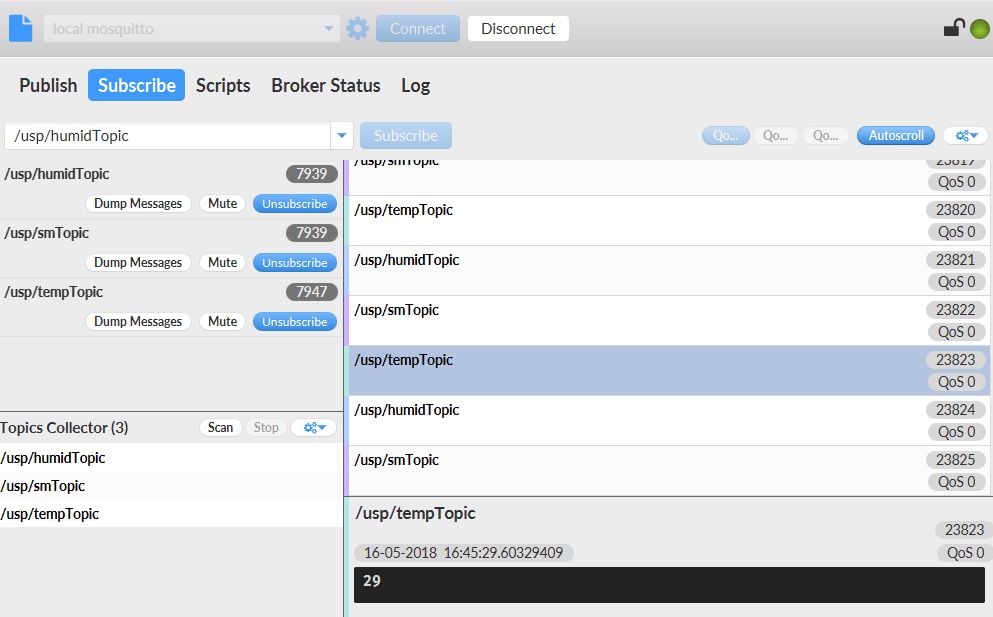}}
 \caption{Sensor data on MQTT broker}
 \label{figure:mqtt}
\end{figure}

\begin{figure}[!ht]
 \centering
 \subfigure[Dashboard home]{\includegraphics[height=4.8cm,width=0.4\columnwidth]{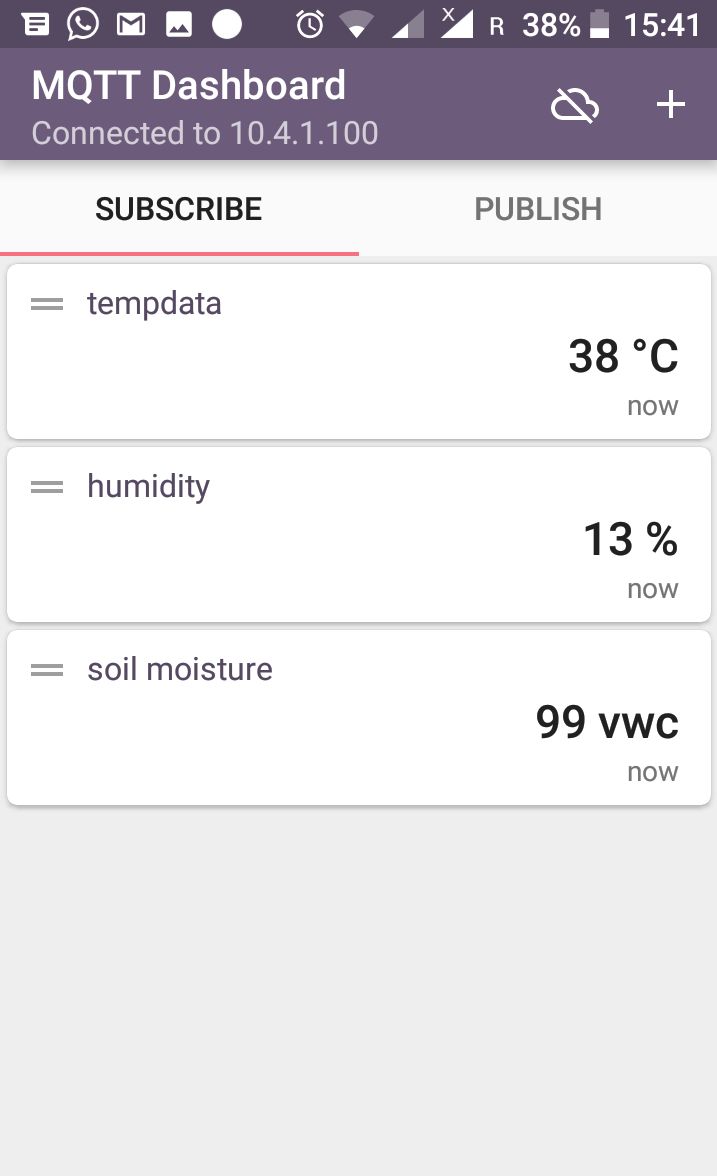}}\hspace{10pt}
 \subfigure[Temperature topic]{\includegraphics[height=4.8cm,width=0.4\columnwidth]{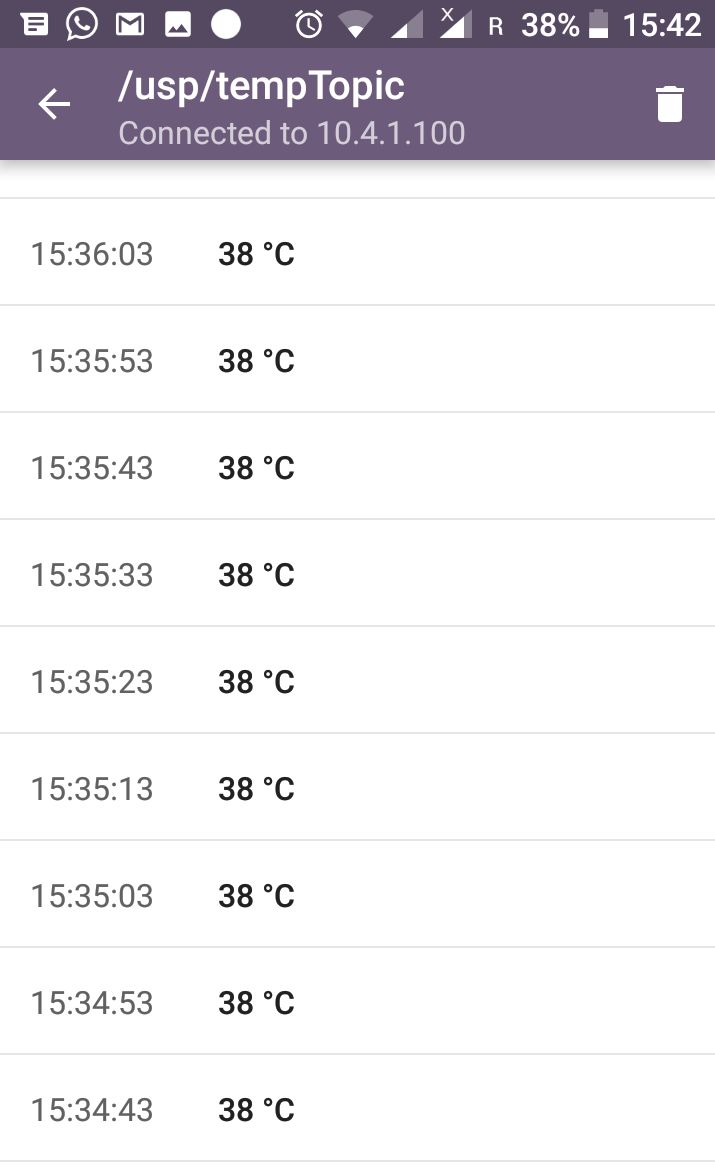}}\hspace{10pt}
 \subfigure[Humidity topic]{\includegraphics[height=4.8cm,width=0.4\columnwidth]{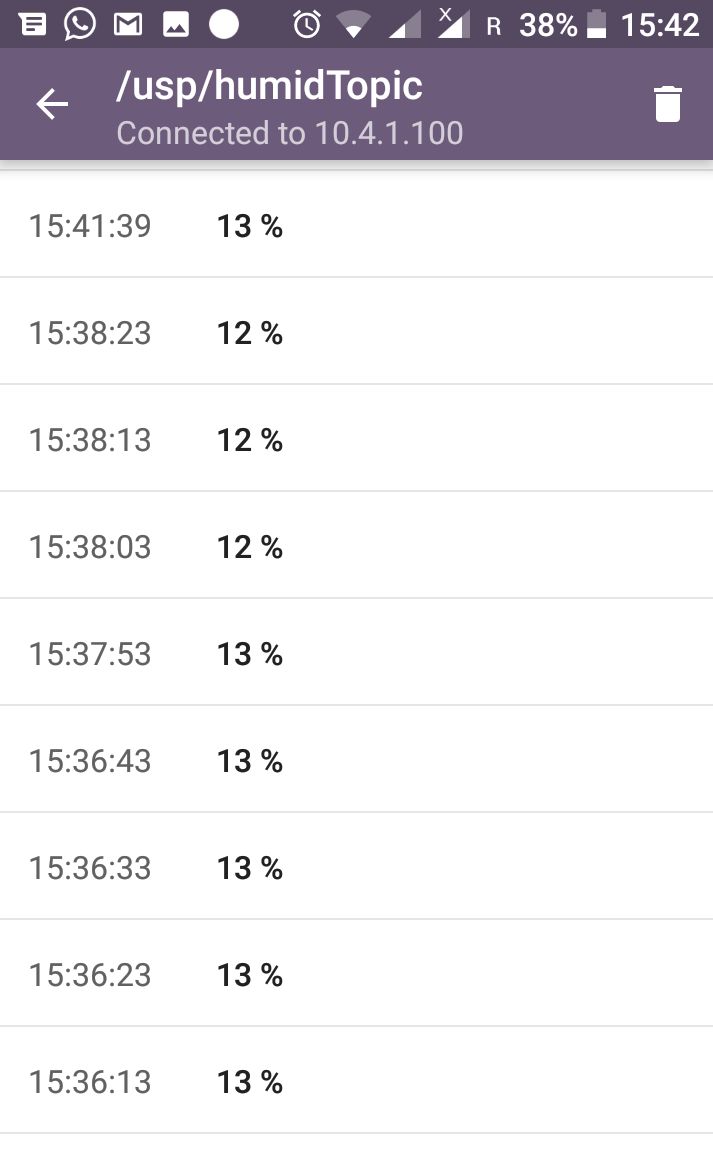}}\hspace{10pt}
 \subfigure[Soil moisture topic]{\includegraphics[height=4.8cm,width=0.4\columnwidth]{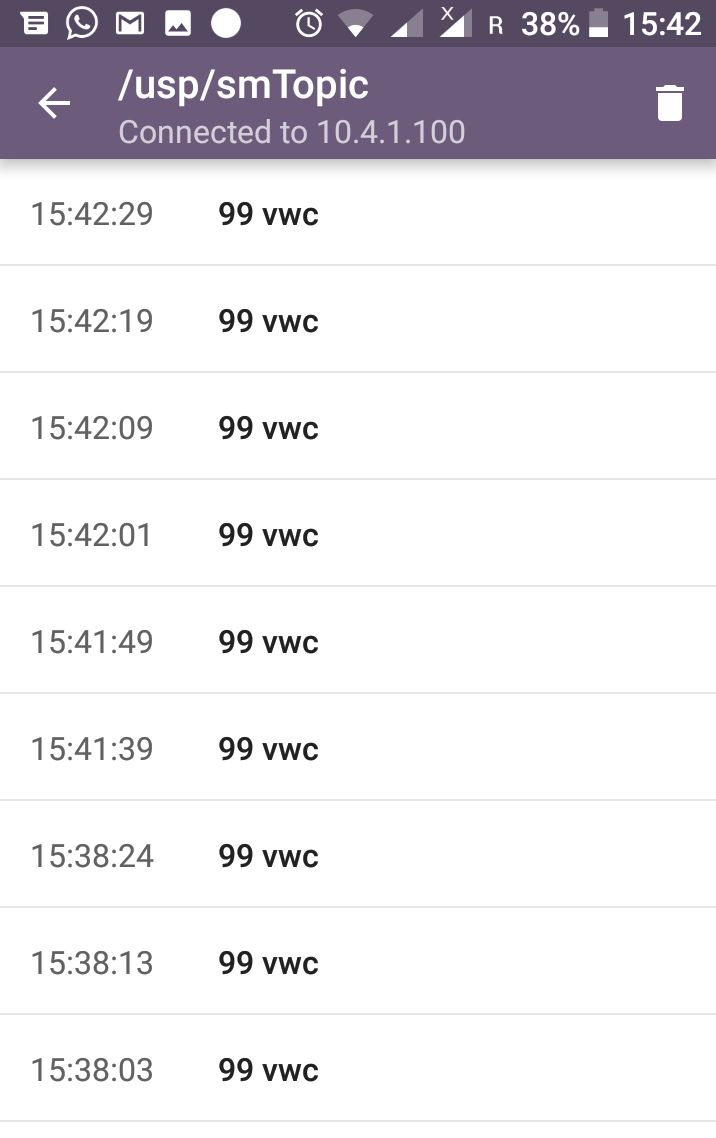}}
 \caption{Sensor data on remote device}
 \label{figure:mqttphone}
\end{figure}

As mentioned earlier, the data fetched via the sensors connected to the device can be remotely monitored using MQTT based server generally referred as broker and Figure \ref{figure:mqtt} shows corresponding results. The Mosquitto broker status is seen from Figure \ref{figure:mqtt} (a), which shows the received and sent data information, number of connections to the broker, published messages information. Figure \ref{figure:mqtt} (b) shows information about the subscribed topics, while the subscribed topicwise sensor data values, timings is seen from Figure \ref{figure:mqtt} (c). This data can also be viewed from an app on a smartphone as shown in Figure \ref{figure:mqttphone}. Here, the MQTT dashboard is first connected to the host Mosquitto broker, which in this case is an IP address rather than a host name. After successful connection is established, it subscribes to topics hosted by the broker i.e. temperature, humidity and soil moisture topics (in this application). This keeps the user updated with the latest data and actuator status. An additional advantage is that if the device fails it will not able updates thus the user gets the knowledge that it has failed/ stopped working and can rectify the issues with it.

The device could also be used for another smart application just by connecting the proper sensors, output tool, providing corresponding inputs and, creating an application setup for analysis.

\section{Conclusion} \label{section:conclusion}
A stand-alone device with, (i) user interaction, html webpage hosting, application specific user requirement, (ii) data sensing, decision making and actuation, (iii) data publish to server capabilities was proposed in this paper. It allows remote access of sensor data, actuator control if needed and, thus eliminates the need of manual operation. For verification purpose a home gardening environment was created using the proposed device along with some farming related sensors. The overall irrigation requirement and schedule obtained for various plants using the device have also been mentioned. It shows flexibility to be used for multiple smart applications like, smart home, organic farming, indoor gardening and organics, etc. Thus with perfect combination of sensors, the device could be used for various other applications as well.

\addcontentsline{toc}{chapter}{REFERENCES}
\bibliographystyle{ieeetr}
\bibliography{reference}
\end{document}